\documentstyle[aps]{revtex}
\begin{document}
\draft
\title{Classical stability and quantum instability
       of black-hole Cauchy horizons}
\author{Dragoljub Markovi\'c}
\address{Department of Physics, University of British Columbia,
         Vancouver, British Columbia, Canada V6T 1Z1}
\author{Eric Poisson\thanks{Present address:
        Department of Physics, Washington University,
        St.~Louis, Missouri 63130.}}
\address{Theoretical Astrophysics,
         California Institute of Technology,
         Pasadena, California 91125}
\date{Draft \today; resubmitted version}
\twocolumn[
\maketitle
\widetext
\begin{abstract}
For a certain region of the parameter space $\{M,e,\Lambda\}$,
the Cauchy horizon of a (charged) black hole residing in de
Sitter space is classically stable to gravitational
perturbations. This implies that, when left to its own devices,
classical theory is unable to retain full
predictive power: the evolution of physical fields beyond
the Cauchy horizon is not uniquely determined by the initial
conditions. In this paper we argue that the Cauchy horizon of
a Reissner-Nordstr\"om-de Sitter black hole must always be
unstable quantum mechanically.
\end{abstract}
]
\narrowtext

It is well known that general relativity admits a well posed
initial value formulation \cite{HE}. This implies that,
given suitable initial data on a spacelike hypersurface
$\Sigma$, the solution to the Einstein equations is uniquely
determined (up to diffeomorphisms) everywhere within the
domain of dependence of the initial surface, $D(\Sigma)$.
For some spacetimes however, $D(\Sigma)$ fails to cover the
whole manifold. The boundary of $D(\Sigma)$ is
known as the Cauchy horizon of the initial surface.
In spacetimes with Cauchy horizons, the field equations lose
their ability to completely determine the future evolution
of the gravitational field; predictability is lost at
the Cauchy horizon.

A particular example of a spacetime with Cauchy horizon, which
shall form the arena of this Letter, is that of a black hole
possessing either electric charge or angular momentum. For
simplicity, we shall restrict our attention to the case of
a non-rotating, spherically symmetric, charged black hole,
as described by the Reissner-Nordstr{\"o}m solution. This
spacetime is well-known to contain a Cauchy horizon
\cite{HE}, which is coincident with the black-hole inner
apparent horizon.

The Cauchy horizon of the Reissner-Nordstr{\"o}m spacetime
is known to be unstable to time-dependent perturbations
\cite{CHinst}. Physically, the instability has to do with
the fact that the causal past of the Cauchy horizon
contains all of the spacetime external to the black hole.
Physical effects occurring outside appear highly
blueshifted as seen by internal observers near the Cauchy
horizon; the blueshift becomes infinite at the Cauchy horizon.

The gravitational field near the Cauchy
horizon of a perturbed Reissner-Nordstr{\"o}m black hole
has been the subject of vigorous investigation
\cite{massinf}. It appears that in the perturbed spacetime,
the Cauchy horizon is replaced by a null curvature
singularity. This singularity is characterized not
by a collapse to zero area, but by an unbounded increase
of the internal mass function \cite{massinf}.

This Cauchy-horizon instability offers an interesting
way out of the loss-of-predictability problem. That the
Cauchy horizon becomes a curvature singularity implies that
the classical evolution of the spacetime cannot proceed
beyond the Cauchy horizon. In this sense, the predictive
power of the theory stays intact. Of course, the classical
theory becomes invalid in the vicinity of the singularity,
where quantum effects are presumably important
\cite{quantum}. The point is that classical theory must
break down {\it before} the Cauchy horizon is encountered.

It is striking that, in the case of black-hole spacetimes
and within the realm of {\it classical} physics,
the Cauchy-horizon instability does not serve as a
{\it universal} mechanism capable of restoring the full
predictive power of the field equations. Indeed,
black-hole spacetimes exist for which the Cauchy horizon is
classically {\it stable}. While this is not generally
possible if the black hole resides in asymptotically-flat
space, stability can be arranged if the black hole is
immersed in de Sitter space.

In this Letter, we show that once quantum physics is
invoked, the full predictive power of general
relativity can be restored: even
when it is classically stable, the Cauchy horizon cannot be
stable quantum mechanically (excluding possibly a
set of measure zero of spacetimes, as we shall explain).
Instability is caused by the divergence
of the renormalized expectation value of the
stress-energy tensor associated with quantized matter
fields.

That quantum effects are needed to prevent a loss
of predictability at the Cauchy horizon is most
intriguing. In this respect,
the physics of black-hole Cauchy horizons is remarkably
similar to that of chronology horizons \cite{Thorne}.

We begin with a brief review of the Reissner-Nordstr\"om-de
Sitter (RNdeS) spacetime \cite{Carter}.

The solution to the Einstein-Maxwell
equations (with cosmological
constant $\Lambda$) representing a charged black hole
in de Sitter space is given by
\begin{equation}
\begin{array}{rcl}
ds^2 &=& -fdv^2 + 2dvdr + r^2(d\theta^2 + \sin^2\!\theta
d\phi^2), \\
f &=& 1 - 2M/r + e^2/r^2 -
{\textstyle \frac{1}{3}} \Lambda r^2.
\end{array}
\label{1}
\end{equation}
Here, $v$ is a null coordinate which is constant
along radial ($d\theta = d\phi =0$), ingoing
($r$ decreasing) null geodesics; $M$ and $e$ are,
respectively, the mass and charge of the black hole.
We use units in which $G=c=1$.

The RNdeS spacetime possesses
three types of horizons. The cosmological horizon
is located at $r=r_c$, and the black-hole outer horizon
at $r=r_e$; the inner horizon ($r=r_i$) is also
a Cauchy horizon for any external
spacelike hypersurface. The roots $r_i < r_e < r_c$
are determined by solving the quartic $f=0$; the fourth
root is unphysical. The surface gravity
of the horizon $r=r_j$ is given by $\kappa_j
= \frac{1}{2} |f'(r_j)|$; here and throughout,
a prime denotes differentiation with respect to the
argument.

We now introduce three observers in the RNdeS
spacetime (Fig.~1). $\cal C$ is a free-falling
observer who crosses the future cosmological horizon;
$r$ increases along $\cal C$'s
world line. ${\cal C}'$ is a static observer, located
just inside the cosmological horizon; along
${\cal C}'$'s world line, $r=r_c(1-\epsilon)$, where
$\epsilon$ is a small, positive constant.
Finally, $\cal I$ is a free-falling
observer who crosses the Cauchy horizon;
$r$ decreases along $\cal I$'s world line.

Mellor and Moss \cite{MellorMoss} have carried out a
classical perturbation analysis for the
Reissner-Nordstr\"om-de
Sitter spacetime, and have shown that a region of
the parameter space $\{M,e,\Lambda\}$ exists
for which the Cauchy horizon is stable.
This region is defined by the inequality
\begin{equation}
\kappa_i \leq \kappa_c,
\label{2}
\end{equation}
which was first written down by Brady and Poisson
\cite{BradyPoisson}. Recently, Chambers and Moss
\cite{ChambersMoss} have shown that Eq.~(\ref{2})
also implies stability for a rotating, charged
black hole residing in de Sitter space.

The condition (\ref{2}) for classical stability can
be motivated simply. The following discussion,
borrowed from Ref.~\cite{BradyPoisson},
will help us understand the differences between
the classical and quantum-mechanical stability
analyses.

We consider, in the fixed RNdeS background, a test
distribution of non-interacting massless particles,
in the continuum limit. The particles originate from
the cosmological region, move radially inward along curves
$v=\mbox{const.}$, and eventually fall into the
black hole. They are
described by the stress-energy tensor
\begin{equation}
T_{\alpha\beta} = \bigl[L(v)/4\pi r^2\bigr]
(\partial_\alpha v) (\partial_\beta v).
\label{3}
\end{equation}
We suppose that $L(v)$ has support all the way
to $v=\infty$. To take into account the coordinate
singularity at the future cosmological horizon, we let
$L(v) = K e^{-2 \kappa_c v}$, where $K$ is a constant,
in the limit $v\to\infty$. A simple calculation then shows
that observer $\cal C$ measures an energy
density $\rho_{\cal C}$ that is everywhere finite and
non-vanishing \cite{foot1}. Observer ${\cal C}'$,
on the other hand, measures an energy density
$\rho_{{\cal C}'}$ that vanishes in the limit
$\epsilon \to 0$; for fixed $\epsilon$,
$\rho_{{\cal C}'} \propto f^{-1} e^{-2\kappa_c v}$.

Next, we consider the influx as measured by
observer $\cal I$. The energy density is now given by
\begin{equation}
\rho_{\cal I} = \bigl(\tilde{E}^2 K/4\pi {r_i}^2\bigr)
e^{-2(\kappa_c-\kappa_i)v},
\label{5}
\end{equation}
where $\tilde{E} \equiv -u_v$ is a constant.
We see that $\rho_{\cal I}$ is
redshifted by the cosmological horizon, and blueshifted
by the Cauchy horizon. It is this (infinite) blueshift
which tends to produce an instability. Nevertheless,
$\rho_{\cal I}$ stays bounded when Eq.~(\ref{2}) is
satisfied, and the Cauchy horizon is then
classically stable.

We now turn to the question of quantum stability.
We assume that quantum fields exist in the RNdeS
spacetime, and we seek to examine the behavior
of $\langle T_{\alpha \beta} \rangle$, the renormalized
expectation value of their stress-energy tensor,
near the Cauchy horizon.

To calculate $\langle T_{\alpha \beta} \rangle$ is
notoriously difficult, even in spherical symmetry
\cite{Anderson}. This is even more true in our case,
because the quantum state cannot be chosen among the
standard ones, such as the Hartle-Hawking or Unruh states
\cite{Hiscock}. We shall therefore consider a simpler
problem, that of quantizing fields in a two-dimensional
version of the RNdeS spacetime. The metric is taken to be
\begin{equation}
ds^2 = -f dudv,
\label{6}
\end{equation}
where the null coordinate $u$ is defined by
$du = dv - 2 f^{-1} dr$. In two dimensions, the
calculation of $\langle T_{ab} \rangle$ can be
carried out explicitly \cite{Davies}.
For simplicity, we shall consider only the
case of a conformally invariant scalar field.
Below we will argue that the conclusions reached
within the two-dimensional model should stay valid
when applied to the four-dimensional spacetime.

We shall express $\langle T_{ab} \rangle$
in the coordinates $(u,v)$. However, we shall
define the quantum state by expanding the
scalar field into positive-frequency modes of
the form $e^{-i\omega \bar{u}}$,
$e^{-i\omega \bar{v}}$, where the transformations
$\bar{u}(u)$ and $\bar{v}(v)$ will be
specified shortly. In this state, and as given in
Ref.~\cite{Balbinot}, the renormalized expectation
value of the stress-energy tensor is (we set
$\hbar = 1$)
\begin{equation}
\langle T_{ab} \rangle =
\theta_{ab} + t_{ab} +
(48\pi)^{-1} R g_{ab}.
\label{7}
\end{equation}
Here, $R$ is the Ricci scalar associated with
the two-dimensional metric; $\theta_{ab}$
is a state-independent object whose
non-vanishing components are
\begin{equation}
\theta_{uu} = -(12 \pi)^{-1} f^{1/2}
{\partial_u}^2 f^{-1/2}
\label{8}
\end{equation}
and $\theta_{vv}$ (which is obtained
by the replacement $u\to v$);
$t_{ab}$ contains the information about
the state and has non-vanishing components
\begin{equation}
t_{uu} = (24 \pi)^{-1} \bigl[
{\textstyle \frac{3}{2}}
\bigl(\bar{u}''/\bar{u}'\bigr)^2 -
\bar{u}'''/\bar{u}' \bigr]
\label{9}
\end{equation}
and $t_{vv}$ (obtained by the
replacements $u \to v$, $\bar{u} \to \bar{v}$).

We need a quantum state that is regular at both
$r=r_c$ and $r=r_e$. Such a state has been constructed
for the
Schwarzschild-de Sitter spacetime by Markovi\'c and
Unruh \cite{Markovic}. It is trivial to generalize
their construction to the RNdeS spacetime. The
state is defined by choosing
$\bar{u}$ to be an affine parameter along the
null geodesic $v=v_0 = \mbox{const.}$,
and $\bar{v}$ an affine parameter along the
null geodesic $u=u_0$.
The null geodesics intersect at a
radius $r_0$ such that $r_e < r_0 < r_c$. These
choices imply
\begin{equation}
\bar{u}' = f[r(v_0-u)], \qquad
\bar{v}' = f[r(v-u_0)],
\label{10}
\end{equation}
where we have indicated that $r$ is a function
of $u$ and $v$, defined implicitly by $2 f^{-1} dr
= dv-du$.

A straightforward calculation, using the equations
listed above, reveals that in the Markovi\'c-Unruh
state, $\langle T_{ab} \rangle$ is given by
\begin{eqnarray}
\langle T_{uu} \rangle &=&
-(48 \pi)^{-1}\bigl[F(u,v)-F(u,v_0)\bigr],
\nonumber \\
\langle T_{uv} \rangle &=&
(48\pi)^{-1} f f'', \label{11} \\
\langle T_{vv} \rangle &=&
-(48\pi)^{-1}\bigl[F(u,v)-F(u_0,v)\bigr],
\nonumber
\end{eqnarray}
where
\begin{equation}
F(u,v) = F[r(v-u)] = {\textstyle \frac{1}{4}}
\bigl( {f'}^2 - 2 f f'' \bigr).
\label{12}
\end{equation}
For our purposes, the detailed functional form of
$F(r)$ is not important. What is essential is that
near a horizon $r=r_j$, $F(r)$ asymptotically
behaves as
\begin{equation}
F(r) = {\kappa_j}^2 - \lambda_j f^2 + O(f^3),
\label{13}
\end{equation}
where $\lambda_j = f'''(r_j)/4 f'(r_j)$ and
$\kappa_j$ is the surface gravity. It
can be checked that
$\langle T_{ab} \rangle$ is regular at $r=r_c$
and $r=r_e$; this can be done by introducing
well-behaved coordinates adapted
to the horizon under consideration.

We now consider measurements made by observer
${\cal C}'$. To remain static, this observer
must be strongly accelerated. In the limit
$f\to 0$, his scalar acceleration is given by
$a \equiv (g_{ab} a^a a^b)^{1/2} = \kappa_c f^{-1/2}$,
where $a^a = u^a_{\,\, ;b} u^b$ is the acceleration
vector. ${\cal C}'$ is therefore immersed in a bath
of thermal radiation (just as an accelerated observer
in Minkowski spacetime would) with Unruh
temperature $T_c = a/2\pi = \kappa_c f^{-1/2}/2\pi$
\cite{Unruh}. (Observer $\cal C$, who is freely falling,
has no indication that this thermal bath exists.) The Unruh
radiation is not all that the static observer sees, for
(outgoing) quanta are also created in the vicinity of
the black-hole past horizon. (These are also seen by
$\cal C$.) These quanta are thermally distributed, and
their locally-measured temperature $T$ varies according
to Tolman's law, $T f^{1/2} = \mbox{const.} = \kappa_e/2\pi$
\cite{Tolman}.

These conclusions are supported by the following result,
obtained with the help of Eqs.~(\ref{11}) and (\ref{13}). At late
times ($u\to\infty$), observer ${\cal C}'$ measures an
energy density given by (up to terms which are finite
when $f \to 0$)
\begin{equation}
\langle \rho_{{\cal C}'} \rangle = (48\pi)^{-1} f^{-1}
\bigl( {\kappa_e}^2 - {\kappa_c}^2 \bigr).
\label{14}
\end{equation}
It is noteworthy that the Unruh radiation contributes
negatively to $\rho_{{\cal C}'}$. This can be understood as
follows \cite{membrane}. Consider an accelerated observer
in Minkowski spacetime, performing measurements on a quantum
field. We suppose that the field is in the true Minkowski
vacuum state. The observer sees a thermal bath at
temperature $a/2\pi$ \cite{Unruh}, and associates to it an
energy density $k(a/2\pi)^n$, where $k$ is a constant,
$a$ the acceleration, and $n$ the dimensionality of
spacetime. The observer also measures the vacuum
polarization created by the quantum field. Because the
renormalized energy density must be precisely zero if
the field is in its true vacuum state, the vacuum
polarization must contribute the negative amount
$-k(a/2\pi)^n$ to the total result. This contribution
does not depend on the quantum state, and is what
appears as the second term to the right of
Eq.~(\ref{14}).

The negative contribution associated
with the vacuum polarization is present in every mode of the
quantum field. Thus, in terms of measurements made by
${\cal C}'$, the vacuum polarization cancels in the
stress-energy tensor the contribution of the thermal flux
emanating from the cosmological horizon. On the other hand,
the thermal flux from the black-hole horizon is {\it not} cancelled
(because of the difference in the horizon temperatures), and
this results in the net (renormalized) flux (\ref{14}), which
is directed toward the future cosmological horizon. We note
that $\cal C$ travels in the same direction as this flux
and, due to redshift, measures a finite energy density. This accounts
for the regularity of the stress-energy tensor at the cosmological
horizon.

This partial cancellation by vacuum polarization explains how
a steady value for $\langle \rho_{{\cal C}'} \rangle$ can be
compatible with a finite value of
$\langle \rho_{\cal C} \rangle$ at
the cosmological horizon. Contrary to the classical case, the
redshift factor $e^{-2\kappa_c v}$ needs not, and does not,
appear in Eq.~(\ref{14}). [Recall the discussion following
Eq.~(\ref{3}).]
{\it This is a key result,} which
has nothing to do with the fact that our expression for
$\rho_{\cal C}$ was
derived for a four-dimensional spacetime, while
Eq.~(\ref{14}) holds in two dimensions. This
result must be attributed to the different physics
associated with the classical and quantum fields.

We are now ready to examine the behavior of the quantum
stress-energy tensor in the vicinity of the Cauchy horizon.
A slight difficulty is that the coordinates $(u,v)$ do not
cover the portion of spacetime inside the black
hole. To remedy this, we transform back to $v$ and $r$,
using $2f^{-1}dr = dv - du$, which
cover both the exterior and interior regions,
up to $v=\infty$. For convenience, we then introduce the
null coordinate $u'$, which is defined only in the interior
region, by the transformation $du' = 2f^{-1} dr - dv$.
It is easy to see that under the combined transformation
$(u,v) \to (u',v)$, Eqs.~(\ref{11})--(\ref{13}) keep the
same form, except that $\langle T_{u'v} \rangle$ comes with
the opposite sign, and that $r$ is now a function of $u'+v$.

We calculate $\langle \rho_{\cal I} \rangle$, the
energy density as measured by observer $\cal I$.
If $\kappa_c \neq \kappa_i$, we find
\begin{equation}
\langle \rho_{\cal I} \rangle =
\bigl( \tilde{E}^2/48\pi \bigr)
\bigl( {\kappa_c}^2 - {\kappa_i}^2 \bigr)
e^{2 \kappa_i v},
\label{16}
\end{equation}
which diverges
in the limit $v\to \infty$.
(The thermal flux, which emanates from
the cosmological horizon and hits the observer
{\it head-on}, is not cancelled by the local vacuum
polarization.) If, on the other hand,
$\kappa_c = \kappa_i$, then $\langle
\rho_{\cal I} \rangle$ is regular at the Cauchy
horizon. We conclude that {\it the Cauchy
horizon of a two-dimensional
RNdeS black hole is quantum-mechanically
unstable} \cite{foot2},
except for the set of measure zero of spacetimes
for which $\kappa_i = \kappa_c$.

We believe that this conclusion also applies to the
four-dimensional spacetime. Indeed, it appears to us
that the fundamental physics of the problem, which is
revealed by the two-dimensional calculation, is robust
and does not depend on the dimensionality of spacetime
(or on the nature of the quantum field --- its spin, conformality,
etc.). Our results can all be intuitively explained in terms
of fundamental processes such as the creation of thermal
quanta near horizons, and the gravitational redshifts
and blueshifts that these quanta undergo. (And once
the contrast between $\rho_{{\cal C}'}$ and
$\langle \rho_{{\cal C}'} \rangle$ is made, it follows
easily from a blueshift argument that, generically,
$\langle \rho_{\cal I} \rangle$ must diverge at the
Cauchy horizon.) These processes take place equally well
in four as in two dimensions \cite{membrane}. Physical
intuition therefore suggests that our two-dimensional
results should be qualitatively valid also in
four dimensions.

We therefore conjecture that, excluding possibly a set
of measure zero of such spacetimes, the Cauchy horizon of
a four-dimensional RNdeS spacetime is always
quantum-mechanically unstable. The mechanism for
instability is the divergence of
$\langle T_{\alpha \beta} \rangle$, the
renormalized expectation value of the stress-energy
tensor associated with quantized matter fields.

Conversations with R.~Balbinot, K.~Thorne, and W.~Unruh were
greatly appreciated. D.M.~was supported by NSERC Grant 580441;
E.P.~was supported by NSF Grant AST 9114925.

\begin{figure}
\caption{Conformal diagram representing a portion of
the RNdeS spacetime. Shown are the cosmological horizons
at $r=r_c$ (future: $v=\infty$; past: $u=-\infty$),
the black-hole outer horizons at $r=r_e$
(future: $u=\infty$; past $v=-\infty$), and
the inner (Cauchy) horizon at $r=r_i$. Also
shown are the asymptotic de Sitter region $r=\infty$,
the timelike singularity $r=0$, and the world lines
of our three observers.}
\end{figure}

\begin{references}
\bibitem{foot1} Here and below, the
measured energy density is given by
$\rho=T_{\alpha\beta} u^\alpha u^\beta$, where
$u^\alpha$ is the observer's four-velocity. A similar
statement holds for the quantum stress-energy tensor.
\bibitem{foot2} Here, we use the phrase ``Cauchy horizon
is unstable'' as a synonym for ``quantum stress-energy
tensor diverges at the Cauchy horizon.'' Because no
field equations have been supplied for the two-dimensional
metric, the question of dynamical stability is not, strictly
speaking, well posed. In four dimensions, however,
the field equations are taken to be the semiclassical
Einstein equations (in which $\langle T_{\alpha\beta} \rangle$
appears on the right-hand side), and the question of
dynamical stability is well posed.
\end{references}
\end{document}